# Harmonic mode locking in a high-Q whispering gallery mode microcavity

Takumi Kato, Tomoya Kobatake, Ryo Suzuki, Zhelun Chen, and Takasumi Tanabe

Department of Electronics and Electrical Engineering, Faculty of Science and

Technology,

Keio University

3-14-1, Hiyoshi, Kohoku-ku, Yokohama, 223-8522, Japan

Tel/Fax: +81-45-566-1730/1529

takasumi@elec.keio.ac.jp

Abstract:

  A numerical and experimental study of the generation of harmonic mode locking in a silica toroid microcavity is presented. We use a generalized mean-field Lugiato-Lefever equation and solve it with the split-step Fourier method. We found that stable harmonic mode locking regime can be accessed when we reduce the input power after strong pumping even when we do not carefully adjust the wavelength detuning. This is due to the bistable nature of the nonlinear cavity system. The experiment agrees well with the numerical analysis, where we obtain low-noise Kerr comb spectrum at low longitudinal mode spacing by gradually reducing the pumping input after strong pumping. This finding clarifies the procedure for generating harmonic mode locking in such high-$Q$ microcavity systems.

  High quality factor ($Q$) microcavities have been investigated intensively, because their high $Q$ and small mode volume allow various applications that require a strong interaction between light and matter [1-4]. One of the attractive applications of a high-$Q$ whispering galley mode (WGM) optical microcavity is frequency comb generation [5,6]. The comb based on a microcavity is called a Kerr comb and its fundamental mechanism was recently revealed [7,8]. A method for mode locking in a microcavity system has been reported both theoretically [9] and experimentally [10-12]. On the other hand, comb formation has been analyzed with modal expansion and a split-step Fourier method [13,14]. These simulations help us to comprehend the transient behavior of the comb formation starting from perturbations; for example we will be able to understand the required input power, optimal input detuning and coupling between a waveguide and a cavity.

In this study, we numerically and experimentally show the generation of harmonic mode locking in a microcavity. In particular, we analyze the generation of harmonic mode locking with a split-step Fourier method, and reveal the procedure required to obtain mode locking with a designed free spectral range (FSR) mode spacing (i.e. 1, 2, and multi-FSRs).

Harmonic mode locking is often studied with a fiber laser to obtain high repetition rate pulses [15]. Complex configurations such as the adjustment of the delay lines and spectral filtering are required for a fiber laser. Furthermore, there is a limitation as regards achieving a high repetition rate, because active mode locking requires a fast electro-optic modulator. In contrast, the study of passive harmonic mode locking in a microcavity has just begun [16,17]. Even though the underlying physics of harmonic mode locking is the same for both a microcavity and a fiber oscillator, the FSR mode spacing $f_\mathrm{R}$ is large and it can satisfy $f_\mathrm{R} > g_\mathrm{FWM}$ in a microcavity, where $g_\mathrm{FWM}$ is the gain bandwidth of the four wave mixing. This characteristic makes the microcavity system unique, because the relationship is usually the opposite for a fiber oscillator. Therefore it is needed to conduct a detailed study to understand the mechanism on how a harmonic mode locking is generated in a microcavity. Here we focus on the multi-FSR comb generation, which will allow us to obtain a simple on-chip optical pulse train source with an ultra-high repetition rate.

To analyze the evolution of the Kerr comb, we use a generalized mean-field Lugiato-Lefever equation (LLE) and solve it with a split-step Fourier method [14,18]. The model is described as,

$$t_R \frac{\partial E}{\partial r} = \left[ -\frac{\alpha}{2} - \frac{\kappa}{2} - i\delta_0 + iL \sum_{k\geq 2} \frac{\beta_k}{k!} \left(\frac{i\partial}{\partial t}\right)^k + i\gamma L |E|^2 \right] E + \sqrt{\kappa} S \qquad (1)$$

where, $t_R$, $\alpha$, $\kappa$, $\delta_0$, $L$, $\beta$, $\gamma$, and $S$ are the round-trip time, intrinsic cavity loss, coupling loss, detuning of the input wavelength, cavity length, dispersion of the cavity, nonlinear coefficient, and the input driving power, respectively. Our fabricated device is a silica toroid microcavity with a radius of 42 μm, and whose cavity length corresponds to the FSR spacing of 800 GHz. A microscope image and the cross-sectional mode profile are shown in the inset of Fig. 1(a). In a microcavity analysis, the exact dispersion $\beta$ is required; so we calculated the material dispersion with the Sellmeier equation and the geometrical dispersion of the resonator with the finite element method (COMSOL Multiphysics). The dispersion for the fabricated structure is shown in Fig. 1(a), and this data is used for the rest of the numerical analysis. We would like to emphasize that our analysis can also be applied to other type of cavities, such as $CaF_2$ and $MgF_2$ WGM cavities, and silicon nitride microrings, by recalculating $\beta$ and use different nonlinear parameters.

First, we show our typical calculation result of the mode locking in a silica toroid microcavity at a wavelength of 1.55 μm. The nonlinear refractive index of silica is $n_2 = 2.2 \times 10^{-20}$ m$^2$W$^{-1}$ and the effective mode volume of the cavity is $A_{\text{eff}} = 2.04$ μm$^2$. Hence $\gamma = 1.13 \times 10^{-8}$ W$^{-1}$μm$^{-1}$. The intrinsic $Q$ of the cavity is $Q_{\text{int}} = 1.0 \times 10^7$, and this value is close to the value we used in the experiment that we describe later. We set the coupling $Q$ at $Q_{\text{couple}} = 1.0 \times 10^7$ and set the detuning of the input laser light at $\delta_0 = 1.0 \times 10^{-4}$ rad. When we fed the cavity with a 20-mW continuous wave (CW) laser light, we obtained a Kerr comb with a spacing of 1-FSR. Figure 1(b) and (c) show the spectral and temporal profiles of the output. It is clear that the generated spectrum is not smooth. Indeed the temporal waveform does not indicate stable pulse generation, and we can see that the cavity is in an unstable state, even though the spectrum appeared as 1-FSR mode-locking. It is because the input is beyond the point of Hopf bifurcation [19-21]. The stable cavity solitons will only exist when the power in the cavity satisfies the condition that suppresses the soliton collision [22].

So we changed the input power to $P_{\text{in}} = 10$ mW; then we obtain four stable pulses in a microcavity. The spectrum is now smooth and shows that the cavity is generating 4-FSR harmonic mode locking. The important fact is that we only modified the input power to switch from unstable 1 to stable 4 FSR Kerr comb generation.

Next we studied the way in which the input power affects the dynamics in the cavity in more detail. For this series of studies we used a cavity with intrinsic and coupling $Q$s of $Q_{\text{int}} = 1.0 \times 10^7$ and $Q_{\text{couple}} = 5.0 \times 10^7$, respectively. The minor diameter of the cavity is 3 μm, and the dispersion is given in Fig. 1(a). It is known that the intra-cavity power exhibits bistable behavior with respect to the input power when the cavity consists of material with a nonlinear refractive index [23]. When we set the normalized detuning at larger than $\sqrt{3}$ times the half width half maximum (HWHM), we can obtain two clear stable states, where $\Delta = 2\delta_0/(\alpha + \kappa)$. Indeed the theoretical curve reveals a nonlinear threshold as shown by the solid line in Fig. 2(a), when the detuning is $\Delta = \sqrt{3}$. This curve is drawn by using an equation where it models the ideal Kerr dispersive bistability in a Fabry-Pérot cavity

The dots in Fig. 2(a) show the intra-cavity power as a function of the input power when $\Delta = \sqrt{3}$, calculated from LLE with the split-step Fourier method. The power in the cavity is recorded after the cavity reaches a steady state. It should be noted that the theoretical curve shown in Fig. 2(a) and (b) covers only single wavelength. This means that nonlinear loss is added when the input power goes into a nonlinear regime. Therefore, the calculated intracavity power travels a different route from the solid line since effective detuning is changed as a result of four-wave-mixing. When we look close to Fig. 2(a), we found that stable 3-FSR harmonic mode locking is only obtained when reduce the input. When we increase the input, the cavity enters an unstable regime before exhibiting mode locking at a power of 7.5 mW.

Figure 2(a) shows that we can obtain 2-FSR mode-locking when we increase the input power if we carefully select the detuning. However, this is not an easy experiment, because the effective detuning easily changes due to the nonlinear loss and thermo-optic effect. Indeed when the detuning is larger than $\Delta = 2$, we cannot obtain 2-FSR mode locking when we increase the power. The result is shown in Fig. 2(b), which is the case when the normalized detuning is $\Delta = 2.2$. But even in this case, it is possible to access 2-FSR mode-locking when we reduce the power after strong pumping; the comb formation changes from 3-FSR to 2-FSR when the input power is reduced.

Our result is consistent to that of a previous report [10], where they show that the repetition rate will change when the input detuning is scanned. However, Figs. 2(a) and (b) provide clear understanding on the relationship between the stable mode-locking and the bistable nature of the cavity system. They show that stable state sits at the bistable regime and usually this regime is accessed only by reducing the input power. In other word, our result shows that the history on how we change the coupled power to the cavity is the key to access to a stable harmonic mode locking. The time and spectrum profile for CW, 3-FSR and 2-FSR mode locking are shown in Fig. 2(c)-(h).

Finally, we describe our experimental demonstration of harmonic mode locking in a silica toroid microcavity. Our setup consists of a tunable laser (Santec TSL-510) that is amplified by an erbium doped fiber amplifier (Pritel LNHP-30) to an output power of up to 1 W. Then the light power is decreased by precisely controlling it with a variable optical attenuator. A polarization controller is placed in front of the cavity. The output from a microcavity is measured with a power meter (Agilent 81634B) and an optical spectrum analyzer (Advantest Q8383). A fast diode (New Focus 1411) and an electrical spectrum analyzer (Advantest R3265A) are used to detect the radio frequency of the amplitude noise of the output Kerr comb. A tapered fiber with diameter of ~1 μm is used as an evanescent coupler to excite the whispering gallery modes in a silica toroid cavity.

Figure 3 shows the measured radio frequency (RF) spectrum at different input powers from a silica toroid microcavity with a $Q$ of $Q_{\text{load}} = 2.0 \times 10^6$. The silica toroid microcavity was fabricated using the method developed by Armani et al. [3]. We successfully observed Kerr comb generation at different FSRs by changing the input power. When the input power was large, the output was 1-FSR as shown in the inset of Fig. 3(a). However, the RF spectrum is noisy, which indicates that the comb is unstable.

On the other hand, when we pump the cavity with lower power, it exhibits stable harmonic mode locking with lower noise in the RF spectrum as shown in Fig. 3(b).

To investigate the above phenomenon in more detail, we performed an experiment where we precisely controlled the input power. We used a cavity with a $Q_{\text{load}} = 2.0 \times 10^6$ that had an FSR of about 800 GHz for this experiment. The coupling is set slightly in an over-coupled regime to compensate for the nonlinear loss. The cavity will be in a critical-coupling condition in a nonlinear regime. The result in Fig. 2 shows that the harmonic mode locking at a lower repetition rate is achieved by reducing the input power. So we started from 200 mW and then reduced the input power. Figure 4(a) is the output spectrum when the input is 200 mW. Although the frequency spacing is 1-FSR, the cavity is in an unstable regime as shown in Fig. 1(b). When we slowly decrease the input power to 183 mW, we obtain a spectrum as shown in Fig. 4(b), where the frequency spacing is now 3-FSR. When we further decrease the input to 58 mW, we obtain 2-FSR spacing as shown in Fig. 4(c). This agrees well with the simulation, where clean harmonic mode locking was obtained simply by reducing the input power.

In summary, we investigated harmonic mode locking in an optical microcavity and found through simulation and experiment that stable mode locking was achieved easily when we reduced the input power after strong pumping. The relationship between the harmonic mode locking and the bistable nature in a microcavity system has been studied, which will help us to reveal the optimal design and procedure with which to obtain an ultra-high repetition rate optical pulse train source for practical applications.

1. K. Vahala, Nature 424, 839 (2003).

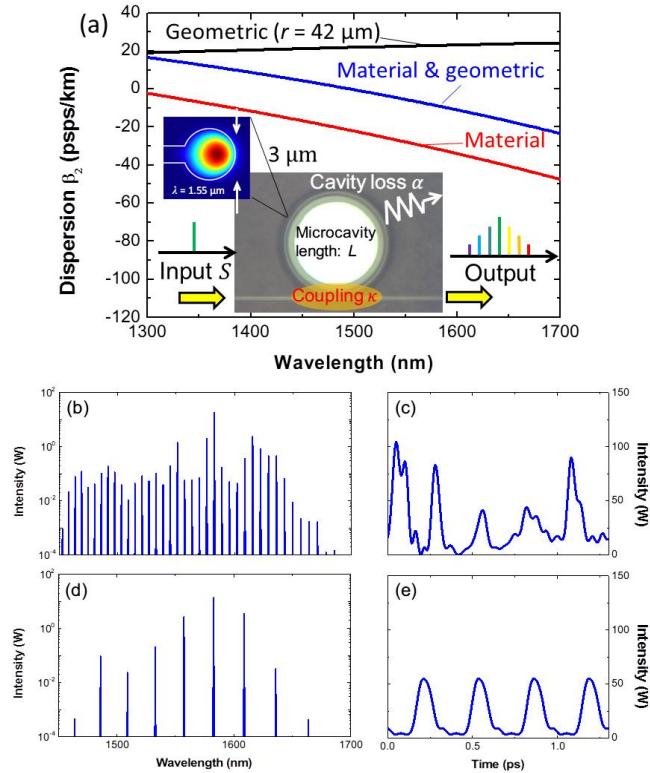

Fig. 1: (a) Material and geometric dispersions of a silica toroid microcavity used for the numerical analysis. The inset illustrates the setup with a microscope image of the silica toroid microcavity coupled with a tapered fiber. A cross-sectional mode profile is also shown. (b) The output spectrum when the input power is 20 mW. (c) Temporal waveform in the cavity for (b). The round trip time of the cavity is 1.3 ps. (d) As (b) but with 10 mW pumping. (e) Temporal waveform in the cavity for (d).

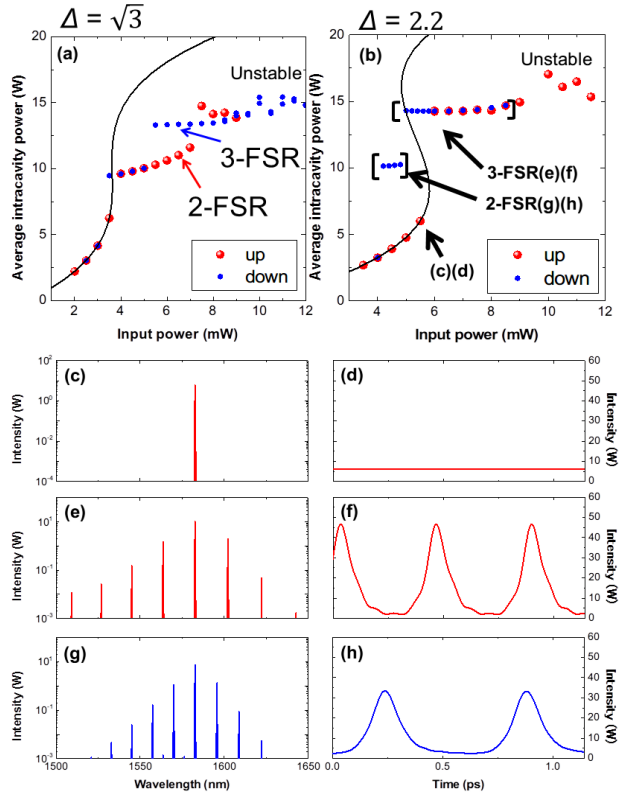

Fig. 2: (a) Calculated average intracavity power as a function of the input power when the normalized detuning of the input laser is $\Delta = \sqrt{3}$ from the resonance of the cavity. (b) As (a) but with detuning $\Delta = 2.2$ (c-h) Spectra and temporal waveforms at different input powers. The corresponding points are shown in (b).

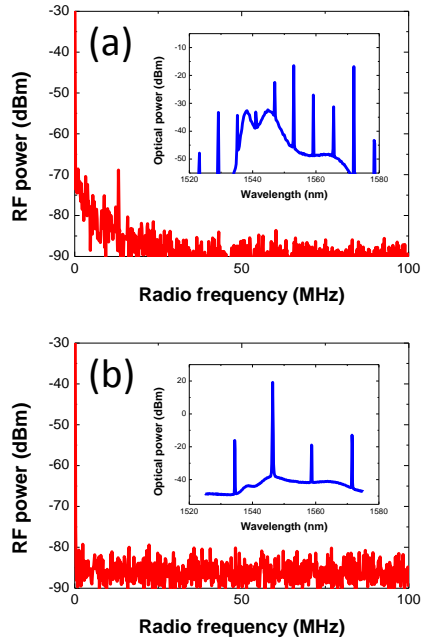

Fig. 3: (a) The experimental RF spectrum of the output from a silica toroid microcavity with strong pumping. The inset is the optical spectrum. (b) As (a) but with weaker optical pumping. The RF frequency is at the noise floor of the measurement apparatus.

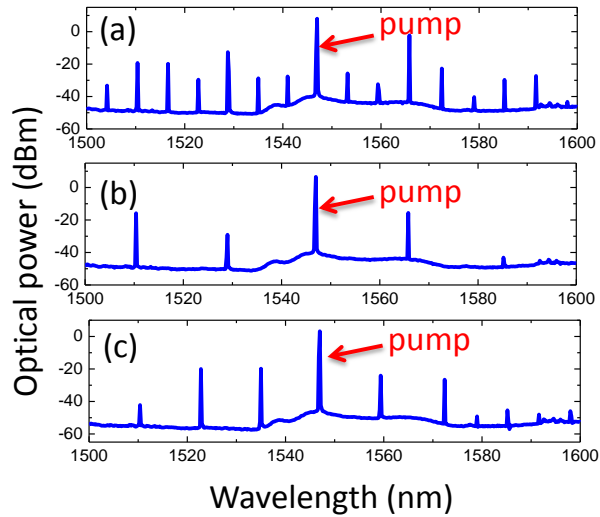

Fig. 4: The output optical spectrum when the cavity is pumped at 200 mW (a), 183.2 mW (b), and 58.6 mW (c). The pumping power was gradually reduced during the experiment.

11